\documentclass[11pt]{article}
\usepackage{geometry}                % See geometry.pdf to learn the layout options. There are lots.
\usepackage[parfill]{parskip}    % Activate to begin paragraphs with an empty line rather than an indent
\usepackage{amssymb}

\usepackage{graphicx}
\usepackage{epstopdf}
\begin{document}

\begin{center}{Preprint of the Institute for Basic Research IBR-EP-39, August 15, 2006}

{submitted for publication}
\vskip1.00cm

{\large {\bf Confirmation of Don Borghi's experiment on the synthesis of neutrons from protons and electrons}}

{\bf Ruggero Maria  Santilli\\ Institute for  Basic Research, 
Box  1577, Palm Harbor,  FL 34682,  U.S.A.\\
  ibr@gte.net, http://www.i-b-r.org, http://www.neutronstructure.org}
\end{center}

\begin{abstract}
Following Rutherford's 1920 historical hypothesis of the  neutron as a compressed hydrogen atom in the core of stars, the laboratory synthesis of the neutron from protons and electrons was claimed in the late 1960 by the Italian priest-physicist Don Carlo Borghi and his associates via a metal chamber containing a partially ionized hydrogen gas at a fraction of $1  bar$ pressure traversed by an electric arc with $5 J$ energy and microwaves with $10^{10}  s^{-1}$ frequency. The experiment remained unverified for decades due to the lack of theoretical understanding of the results. In this note we report various measurements showing that, under certain conditions, electric arcs within a hydrogen gas produce neutral, hadron-size entities that are absorbed by stable nuclei and subsequently result in the release of detectable neutrons, thus confirming Don Borghi's experiment. The possibility that said entities are neutrons is discussed jointly with other alternatives. Due to their simplicity, a primary scope of this note is to stimulate the independent re-run of the tests as conducted or in suitable alternative forms.
\end{abstract}

In 1920, Rutherford [1a] submitted the
hypothesis that hydrogen atoms in the
core of stars are compressed into a new particle having
the size of the proton that he called the {\it neutron,}
according to the synthesis 
$$
p^+, e^- \rightarrow n.
\eqno(1)
$$
The existence of the neutron was confirmed in 1932 by
Chadwick [1b] and its main features identified.
Pauli [1c] noted that the
spin 1/2 of the neutron
cannot be represented via a quantum state of two particles
each having spin 1/2, and conjectured the possible
emission of a new neutral massless particle with spin
1/2. Fermi [1d] adopted Pauli's conjecture, coined the
name {\it neutrino} (meaning in Italian ``little
neutron") with symbol $\nu$, and
developed the theory of {\it weak interactions}
according to the familiar relations 
$$
p^+ + e^- \rightarrow n + \nu,
\eqno(2)
$$
which theory was more recently developed into the so-called {\it standard model} [1e].

Despite historical advances along the latter lines, Rutherford's legacy of synthesizing the neutron from protons and electrons, did not remain ignored. In fact,  the Italian priest-physicist Don Carlo Borghi in collaboration with experimentalists from the University of Recife, Brazil [2a], claimed in the late 1960s to have achieved the laboratory synthesis of neutrons from protons and electrons. 

Don  Borghi's experiment is reviewed in detail in volume [2b] (see in particular L. Daddi [2c] and P. Giubbilini [2d]). We here merely recall that the experiment was conducted via a cylindrical metallic chamber (called "klystron") filled up with a partially ionized hydrogen gas at a fraction of $1 bar$ pressure, traversed by an electric arc with about $500  V$ and $10  mA$ as well as by microwaves with $10^{10} s^{-1}$ frequency. In the cylindrical exterior of the chamber the experimentalists placed various materials suitable to become radioactive when subjected to a neutron flux (such as gold, silver and others).  Following exposures of the order of weeks, the experimentalists reported nuclear transmutations due to a claimed neutron flux of the order of $10^4  cps$, apparently confirmed by beta emissions not present in the original material.

Don Borghi's claim remained ignored for decades due to its incompatibility with established knowledge. 
 Even though not widely spoken in the technical literature, {\it synthesis (2) violates the conservation of  energy unless the proton and the electron have a minimum energy of $0.78  MeV,$ in which case no energy is left for the neutrino.} This is due to the fact that   the sum of the rest energies of the proton and the electron ($938.78   MeV$) is $0.78  MeV$ {\it smaller} than the neutron rest energy ($939.56  MeV$). In the event protons  and electrons do have the missing energy, quantum mechanics predicts that their cross section is so small (about $10^{-20}  barn$) to prevent any possible synthesis. For these and additional unresolved theoretical aspects of Don Borghi's experiment we refer the interested reader to L. Daddi [2c] and P. Giubbilini [2c].

%%%%%%%%%%%%%%%%%%%%%%%%%%

%Figure1

\begin{figure}[htbp] %  figure placement: here, top, bottom, or page
   \centering
   \includegraphics[width=4in]{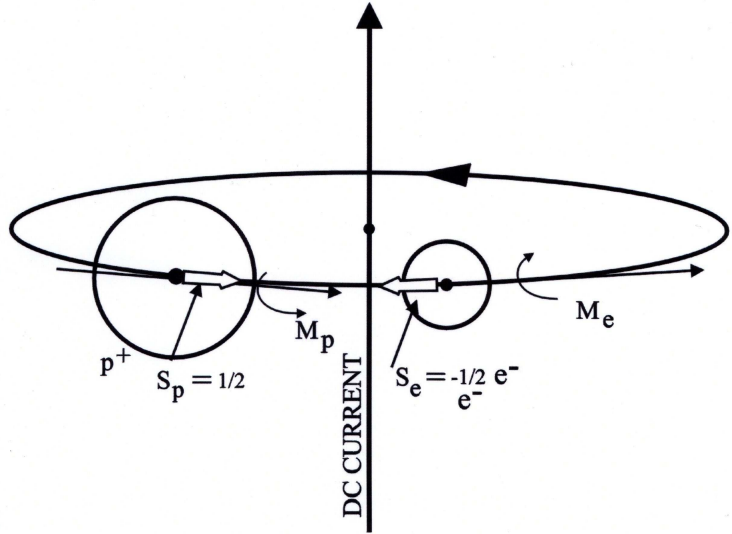} 
 \centering   \caption{{\it A schematic view of the alignment of protons and electrons along a magnetic force line of a DC electric arc. }}
\end{figure}

%%%%%%%%%%%%%%%%%%%%%%%%%%%%%%%%%%%%%%%%%%%%%

In this note we report various measurements  showing that, under conditions and processes identified below, {\it electric arcs within a hydrogen gas cause processes resulting in the  synthesis of neutrons, thus confirming Don Borghi's experiment;} we recommend the independent re-run of our measurements due to their simplicity; and we propose alternative tests. In essence, Don Borghi and his associates used the electric arc for the sole purpose of maintaining the hydrogen gas inside the chamber at least partially ionized [2a]. The measurements presented herein indicate that neutrons originate from processes caused precisely by the arc.

It is important to show that the objections against Don Borghi's experiment are essentially due to a rather widespread belief that novelty beyond quantum mechanics cannot exist. To begin, {\it Don Borghi's experiment does not violate the principle of conservation of the energy} because its arc has about $5  J$ energy, $ 1  eV = 1.60\times 10^{-19}  J$, and $0.78  MeV = 1.24\times 10^{-13} J$. Consequently, the use of $10^{-9}  J$ of the arc energy are sufficient to provide the energy needed for the claimed neutron flux of  $10^4 cps$, the available $5  J$ being essentially dissipated.

Also, the geometry of the electric arc, illustrated in Fig. 1, is quite conducive to the synthesis of the neutron, since it aligns protons and electrons with their magnetic moment along the tangent of the local magnetic force, thus resulting in the pairing of protons and electrons under very large Coulomb attractions (due to the short distances) for both opposite charges and opposite magnetic polarities.  A neutral state of the type conceptually depicted in Fig. 2 is then possible, thus rendering plausible the synthesis of the neutron either under suitable processes or under a suitable "trigger" [3a], such as the microwaves used in Don Borghi's experiment, the fast pressure surge used by the author as reported below, or the compression that appears to be caused naturally by electric arcs with high voltage, as suspected in lighting and related thunder. The remaining objections against Don Borghi's experiment equally emerge as not withstanding serious scientific inquiries.
.

%%%%%%%%%%%%%%%%%%%%%%%%%%

%Figure2

\begin{figure}[htbp] %  figure placement: here, top, bottom, or page
   \centering
   \includegraphics[width=4in]{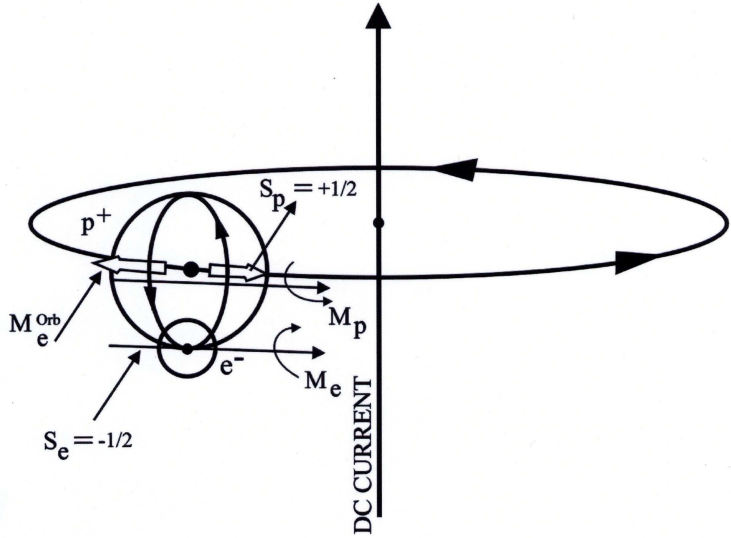} 
 \centering   \caption{{\it A schematic view of the collapse of the electron on thew proton structure as expected from attractive Coulomb forces due to opposing charges and magnetic moments. }}
\end{figure}

%%%%%%%%%%%%%%%%%%%%%%%%%%%%%%%%%%%%%%%%%%%%%

In view of the lack of serious scientific objections, the author decided to conduct measurements of neutrons expected from a chamber containing hydrogen solely traversed by an electric arc. All tests were conducted at the lab of the Institute for Basic Research in Palm Harbor, Florida. The first measurements were conducted via a sealed  cylindrical chamber of about $ 6"$ diameter and $12"$  height made of commercially available, transparent, PolyVinyl Chloride (PVC)  housing along its symmetry axis a pair of tungsten electrodes of $0.250"  OD$ and $1"$ length fastened to the tip of  $0.250"  OD$ conducting rods protruding through seals out of the top and bottom of the chamber for electrical connections.

The electrodes gap was controllable by sliding the top conducting rod through the seal of the flange. The PVC was selected to be transparent so as to allow a visual detection of the arc. Following flushing of air, the chamber was filled up with commercial grade hydrogen at $25  psi$ pressure. For power unit we used an ordinary AC-DC welder, a Miller  Sincrowave 300. Radiation counts were done via  the sensitive photon-neutron detector model $PM1703GN$, s. n. $52777$, purchased from Berkeley Nucleonics and produced by Polimaster, Inc., with  sonic and vibration alarms as well as memory for all readings. 

The first test was done on Friday, July 27, 2006, at 2 p.m. We first used detector  $PM1703GN$ to verify the  background radiations solely consisting of photon counts of $5-7 \mu R/h$  without any neutron count; we delivered a DC electric arc at $27 V$  and $10 A$ (namely with about $55$ times the energy of the arc used in Don Borghi's test), with about $0.125 "$ gap for about $3  s$; we waited until the incandescence of the electrodes was no longer visible (to prevent the override of the photon over the neutron count); and then placed detector $PM1703GN$ against the PVC cylinder. This resulted in the detection of photons at the rate of $10-15 \mu R/h$ expected from the residual excitation of the tips of the electrodes, but no neutron count at all. 

However, about $3 m$ following the test, the detector entered into sonic and vibration alarms, specifically, for neutron detection off the instrument maximum of  $99  cps$ while no anomalous  photon emission was measured. The detector was moved outside the laboratory and the neutron counts returned to zero. The detector was then returned into the laboratory and we were surprised to see it entering into sonic and vibrational alarms at about $5'$ away from the arc chamber, again, with the neutron count off scale without appreciable detection of photons, at which point the laboratory was evacuated for safety. After waiting for $30 m$ (double neutron's lifetime), we were surprised to see detector $PM1703GN$ go off scale again in neutron counts at a distance of $10'$ from the experimental set up. Being in late Friday afternoon, the lab was closed for the day. 

 Inspection of the lab the following morning indicated no neutron detection in the general area, but detector $PM1703GN$ showed clear neutron counts when placed next to the PVC wall. The same detections persisted for two subsequent days until the hydrogen was flushed out of the chamber.  
 
 The test was repeated the afternoon of the following Friday with the welder operating in AC mode at $30 V$ and $30 A$ plus a transformer that allowed to deliver an arc with $700 V$ and $1.2 A$ for $5 s$ with a gap of about $0.375"$. We waited again  until the incandescence of the tips was extinguished and placed detector $PM1703GN$ near the cylindrical PVC wall, resulting in sonic and vibrational alarms much sooner and definitely bigger than those of the first test, thus requiring, again, the evacuation of the lab. 
 
 A third test was done the following Wednesday in a rectangular, transparent, PVC chamber $3"  \times 3"  \times 6"$ filled up with commercial grade hydrogen at atmospheric pressure and temperature traversed by a $2"$ long intermitted electric arc powered by a standard Whimshurst electrostatic generator. Repeated tests produced no neutron detection.  

However, the repetition of the test the following morning caused an unexpected implosion due to the contamination of the chamber with air and the resulting $H-O$ combustion triggered by the spark.  Despite the rudimentary nature of the equipment, this implosion caused, by far, the biggest detection of neutrons via continuous alarms due to off-scale $cps$ without any appreciable photon detection, as confirmed and documented by the print-outs of the $PM1703GN$ detector.
 
 Identical re-runs of the test produced exactly the same results. Detector $PM1703GN$ was returned to the manufacturer for control; it was verified to operate properly; and the printout of all readings stored in its memory was released confirming the measurements reported above. All additional tests were then halted pending the implementary of regulatory procedures.

It should be reported that no meaningful neutron counts were detected in using various other gases, or for electric arcs submerged within liquids. Therefore, the measurements herein reported, in the way conducted, appear to be specifically and solely applicable for electric arcs in {\it gaseous hydrogen}.

In summary, under sufficient power or a suitable "trigger" under  insufficient power,  electric arcs within a  hydrogen gas produce "entities" that 1) are not hydrogen atoms (because in that case the $PM1703GN$ counter would show no detection); 2) have dimensions of the order of those of all hadrons (otherwise, again, the counter would not detect them); 3) are necessarily neutral (to move freely through the PVC walls); 4) are essentially stable for hadron standards (more accurate data being grossly premature at this writing); and 5) remain initially confined within the arc chamber under steady conditions to slowly exit, except for the case of production under implosion causing rapid propagation.

The theoretical interpretation of these findings will be presented elsewhere [3b]. At this moment we limit ourselves to indicate the main possible interpretations of the above findings.

{\bf First interpretation:} this possibility is along Rutherford's historical legacy (1), namely, that  the "entities" produced by the electric arc, rather than being neutrons, are a new bound state of protons and electrons at short distances. In fact,  the entity depicted in Fig. 2 is a {\it Boson} with spin $0$ here tentatively called {\it arcogen} in the Greek meaning of being "arc generated"), while the neutron as notoriously spin ${1\over 2}$. Hence, the arcogen rest energy is expected to be the sum of the rest energies of proton and electron ($938.78   MeV$) less a large Coulomb binding energy, thus being considerable {\it smaller} than the neutron rest energy ($939.56  MeV$).

This alternative is due to the possibilities that: i)  ordinary stable nuclei absorb the arcogens; ii)  said nuclei, rather than the arc, synthesize neutrons according to the standard model, i.e., reaction (2); and iii) ordinary neutrons are then released by the nuclei as part of their radioactive process. In this case counter $PM1703GN$ does indeed detects neutrons, but they are of nuclear origin, rather than synthesized by the electric arc.  

 In fact, Don Borghi was rather cautious in calling "neutrons" the entities produced by his klystron [2a]. The same caution  has been voiced by Daddi [2c] and others [2b]. Also. it should be noted that  new bound states of protons and electrons below the Bohr's ground state of the hydrogen have been predicted by various authors (see, e.g., Mills [2e]) and appear to be necessary in any case for a serious interpretation of the Sun spectral emission.

{\bf Second interpretation:} due to the lack of serious scientific objections, it is also possible that electric arcs within a hydrogen gas directly synthesize the neutron according to the standard reaction (2). In this case the energy needed for the neutron synthesis is provided by the arc, while the conservation of angular momentum is salvaged by the creation and release of an electron neutrino. According to this interpretation, following their synthesis, the neutrons slowly migrate, are absorbed by the stable nuclei of the lab environment and then released as part of natural radioactive processes.

{\bf Third interpretation:} due to intriguing cosmological implications, we should also mention the possibility that neutrons are indeed synthesized by the electric arc, but both the missing energy and angular momentum are provided by the ether (empty space conceived as a universal substratum  of energy) via a neutral entity here tentatively called {\it neutron etherino} (see [3b] for muon, tau and  other particle syntheses) and denoted with the symbol $a$ (from the Latin aether). The (neutron) etherino is, therefore, conjectured to have: zero mass and charge,  the energy of at least $0.78  MeV$, and spin ${1\over 2}$. In this case, rather than reaction (2), we would have the alternative form
$$
p^+ + a + e^- \rightarrow n.
\eqno(3)
$$

The etherino hypothesis should not be dismissed lightly prior to due study because the idea of continuous creation of matter in the universe has been repeatedly voiced. The view here submitted  is that, in the event real, the most plausible mechanism of continuous creation is given precisely by the synthesis of the neutron inside stars. 

 Note that the alternative formulation $p^+ + \bar{\nu} + e^- \rightarrow n$ has no scientific meaning due to the impossibility for proton and electrons to bound to or absorb antineutrinos. Also, rather than  being in conflict  with current claims of neutrino detection, synthesis (3) could be of help in resolving at least some of  the controversies on neutrinos. For instance, according to the standard model, a neutral entity carrying energy in our spacetime, such as the neutrino, can traverse very large hyperdense media such as entire stars without any collision. Such an idea is repugnant to the scientific reason of various physicists, including the author. The etherino hypothesis automatically resolves this issue (since no energy would move in our spacetime), while not necessarily being in conflict with available neutrino experiments and actually having the potential of initiating possible new communications through the ether. 
 
 After all, stars initiate their life as being solely composed of hydrogen and quantitative studies beyond the level of personal beliefs on the availability of the necessary energy for the synthesis of the neutron have not been conducted to the author's best knowledge. There are also serious doubt as to whether  the immense energy needed in supernova explosion can be truly explained via nuclear syntheses alone and the rather limited level of $20$-th century knowledge. Consequently, it is perhaps time for the scientific community to admit the possibility that stars contain mechanisms for extracting energy from space.

A primary objective of this note is to stimulating the independent re-run of the tests by interested colleagues due to their simplicity, low cost and significance. A much needed alternative test is one permitting the measurement of the energy needed for the neutron synthesis, something not possible in Don Borghi's test.

%%%%%%%%%%%%%%%%%%%%%%%%%%

%Figure2

\begin{figure}[htbp] %  figure placement: here, top, bottom, or page
   \centering
   \includegraphics[width=4in]{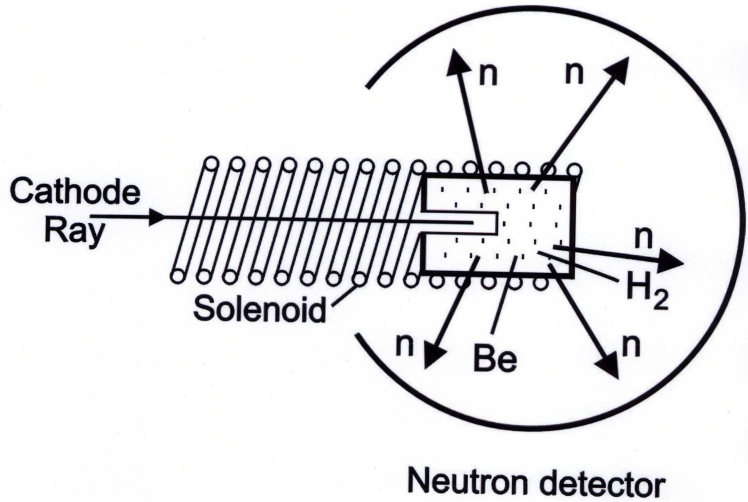} 
 \centering   \caption{{\it A schematic view of the proposed alternative test on the possible laboratory synthesis of the neutron. }}
\end{figure}

%%%%%%%%%%%%%%%%%%%%%%%%%%%%%%%%%%%%%%%%%%%%%

We here propose the conduction of the simple experiment of Fig. 3 essentially consisting of: A)  a cylinder of beryllium saturated with hydrogen and kept at low temperature (so as to minimize thermal motion of their proton); B) a coherent electron beam with a measurable energy smaller, equal or bigger than $0.78  MeV$; C) a magnetic field encompassing both the beryllium bar and the electron beam in order to achieve the axial coupling of Fig. 1 [3b]; D) means to realize a "trigger", such as Don Borghi's high frequency microwaves or the fast pressure surge of this note; and E) a suitable neutron detector.

The resolutory character of the proposed test should be noted. In fact, the test would establish the existence of the (electron) neutrino (only) via the measurement of the electron beam energy if {\it bigger} than $0.78 MeV$ after subtracting the kinetic energy of the neutrons. However, the possible detection of neutrons synthesized for energies {\it smaller} than $0.78 MeV$ would disprove the existence of the neutrino in favor of the etherino hypothesis and the continuous creation in the universe. The case of the possible detection of  neutrons synthesized  at the threshold electron beam energy of $0.78  MeV$ would perhaps be even more intriguing because it would establish that the sole exchange between matter and the ether is given by spin. 

Whatever the alternative that will result to be applicable among the three possibilities submitted above, the measurements presented in this note do provide a direct experimental verification of Don Borghi's experiment. The measurements also support the long suspected capability of lighting to synthesize natural elements (a condition apparently necessary for a serious interpretation of thunder due to the numerical failure of standard interpretations), the synthesis of the neutron being the first of the chain.  Finally, the reported measurements provide encouraging support for the recently proposed "Controlled Intermediate Nuclear Fusion"  [3a] that is precisely based on electric arcs, the control being of easy industrial realization via the control of power, pressure and rigger.

In closing, it is recommendable to note that, irrespective of whether the neutron is synthesized by stars, nuclei or arcs, and irrespective of whether according to the neutrino hypothesis (2) or the etherino hypothesis (3), the synthesis of the neutrons from protons and electrons signals the limit of exact applicability of quantum mechanics. This is due to the fact that quantum mechanics can only describe bound states with the familiar {\it negative} binding energy as occurring in nuclei, atoms and molecules, while the synthesis of neutrons (as well as all hadrons) requires a {\it positive} binding energy as in Eqs. (2) and (3). Under this condition, the Schroedinger equation no longer provide physical results, as the reader is encourage to verify by attempting the solution of any conventional bound state in which the binding energy is turned from a negative to a positive value.

This occurrence required the necessary existing from the class of unitary equivalence of quantum mechanics via the proposal by this author back in 1978 [3c] to construct a nonunitary covering under the name of {\it hadronic mechanics.} The new mechanics was subsequently developed, verified and applied by various authors (see [3d-3f] and quoted references), and achieved the numerically exact and invariant, nonrelativistic [3g]  and relativistic [3h] representation of {\it all} characteristics of the neutron as a generalized bound state of a proton and an electron [3i]. 

The generalized structure equation for the neutron is simply achieved via a suitable nonunitary image of the corresponding equation for the hydrogen atoms, in which nonunitarity represents contact non-Hamiltonian (thus necessarily nonunitary) interactions caused by the deep penetration of the electron wavepacket within the hyperdense medium inside the proton. A similar nonunitary image of the positronium permits the numerically exact and invariant representation of {it all} characteristics of the $\pi^o$ meson as a generalized bound state of one electron and one positron, fully achieved since the original proposal of 1978 (see [3c], Section 5). Similar structure models follow for all unstable hadrons with physical constituents released free, usually in the spontaneous decays with the lowest mode.  

 Compatibility with $S(3)$-color theories is possible under their re-interpretation as solely providing the final {\it Mendeleev-type classification} of hadrons. According to this view, quarks  remain necessary for the elaboration of the theories, but only as mere mathematical quantities (mathematical representation of a mathematical symmetry in a mathematical, complex-valued unitary space with no possible connection with our spacetime), since quark are afflicted by a litany of unresolved inconsistencies when interpreted as physical particles in our spacetime (lack of gravity, since the latter is solely definable in our spacetime as well known since Einstein; lack of inertia,  since the latter is solely definable via the Poincare' symmetry that is inapplicable to quarks as well known, etc.).

All in all, rather than having reached the end of knowledge with quantum mechanics,  research in the synthesis and structure of the neutron may stimulate a new scientific renaissance with theoretical experimental and industrial advances perhaps beyond our imagination at this writing.

{\bf Acknowledgments.} The author would like to thank: Robert Rao and Jim McQuaid of Berkeley Nucleonics and Arif  Mamedov of Polimaster, Inc.,  for invaluable technical assistance on neutron detection; L. Daddi, P. Giubbilini and U. Mastromatteo for deep insights in Dob Borghi's experiment;' and H. E.  Wilhelm, C. Whitney, G. Mileto and others for invariable critical comments.


\begin{thebibliography}{5}

\bibitem{1}
H.~Rutherford Proc. Roy. Soc. A {\bf 97}, 374 (1920) [1a].
J.~Chadwick Proc. Roy. Soc. A {\bf 136}, 692 (1932) [1b].
W.~Pauli, Handbruch der Physik Vol. 24, Springer-Verlag, Berlin (1933)
[1c].
E.~Fermim {\it Nuclear Physics,} University of Chicago Press (1949)
[1d].
M.~Kaku, {\it Quantum Field Theory,} Oxford University Press, New York
(1993) [1e].

\bibitem{2}
C. Borghi, C. Giorio and A. Dall'Olio, "Experimental. evidence on the emission of neutrons from a cold hydrogen plasma," communications of CENUFPE number 8 (1969) and 25 (1971), reprinted in the (Russian) Phys. Atomic Nuclei {\bf 56}, 205 (1993) [2a]. 
{\it Collected Papers on Don Borghi's Experiment,} R. M. Santilli, editor, International Academic Press (in press) [2b]. L. Daddi, contributed paper in {\it Collected Papers on Don Borghi's Experiment} [2c]. P. Giubbilini contributed paper in {\it Collected Papers on Don Borghi's Experiment} [2d]. R. Mills, Int. J. Hydrogen Energy {\bf 25}, 1171 (2000) [2e].

\bibitem{3}
R. M. Santilli, "The novel 'Controlled Intermediate Nuclear Fusion' and its possible industrial realization as predicted by hadronic mechanics and chemistry,"  J. Applied Sciences (in press), also available at
http://arxiv.org/abs/physics/0602125 [3a]; 
"Neutrinos or etherino?", contributed paper to IARD 2006, to appear [3b];
.Hadronic J. {\bf 1}, 224, 574 and 1267 (1978) [3c]; 
Rendiconti  Circolo  Matematico  di
Palermo, Supplemento {\bf {42}}, 7 (1996) [3d]; 
Found. Phys {\bf 27}, 635 (1997) [3e]; 
 Acta Appl. Math. {\bf 50}, 177 (1998) [3f]; 
 Hadronic J. {\bf 13}, 513 (1990) [3g]; 
JJINR Comm. E4-93-352 (1993) published in Chinese J. Syst. Eng. and Electr. {\bf 6}, 177 (1995) [3h]; 
{\it Elements of Hadronic Mechanics,} Vol. I (1995), Vol. II (1995), Vol. III (in press), see also
http://www.i-b-r.org/Hadronic-Mechanics.htm [3i].


\end{thebibliography}
\end{document}